\documentstyle[multicol,prd,aps,graphicx]{revtex}
\begin{document}
\draft

\title{ 
\vglue -0.5cm
\hfill{\small IFT-P.089/2002} \\
\hfill{\small hep-ph/0211107} \\
\hfill{\small November 2002}\\
\vglue 0.5cm
Naturally light invisible axion in models with large local discrete symmetries
} 

\author{Alex G. Dias$^1$, V. Pleitez$^2$ and M. D. Tonasse$^3$}
\address{$^1$Instituto de F\'\i sica, Universidade de S\~ao Paulo, \\
C. P. 66.318, 05315-970\\ S\~ao Paulo, SP\\ Brazil} 
\address{$^2$Instituto de F\'\i sica Te\'orica, Universidade Estadual Paulista,\\
Rua Pamplona 145, \\
01405-900 S\~ao Paulo, SP \\Brazil} 
 \address{$^3$Instituto Tecnol\'ogico de Aeron\'autica, Centro T\'ecnico  
Aeroespacial\\ Pra\c ca Marechal do Ar Eduardo Gomes 50, 12228-901 \\
S\~ao Jos\'e 
dos Campos, SP\\ Brazil}  

\date{\today}
\maketitle

\begin{abstract}
We show that by introducing appropriate local $Z_N(N\geq13)$ symmetries in
electroweak models it is possible to implement an automatic Peccei-Quinn
symmetry keeping at the same time the axion protected against gravitational
effects.  Although we consider here only an extension of the standard model and
a particular 3-3-1 model, the strategy can be used in any kind of electroweak
model. An interesting feature of this 3-3-1 model is that if: {\it i)} we add
right-handed neutrinos, {\it ii)} the conservation of the total lepton number,
and {\it iii)} a $Z_2$ symmetry, the $Z_{13}$ and the chiral Peccei-Quinn
$U(1)_{\rm PQ}$ are both accidental symmetries in the sense that they are not
imposed on the Lagrangian but they are just the consequence of the particle
content of the model, its gauge invariance, renormalizability and Lorentz
invariance. In  addition, this model has no domain wall problem.   

\end{abstract}
\pacs{PACS numbers: 14.80.Mz; 12.60.Fr; 11.30.Er }


\begin{multicols}{2}
\narrowtext
\section{Introduction}
\label{sec:intro}

It is well known that an elegant way to solve the strong CP problem is by 
introducing a chiral $U(1)_{\rm PQ}$~\cite{pq} symmetry which also implies
the existence of a pseudo-Goldstone boson--the axion~\cite{axion}. This particle
becomes an interesting candidate for dark matter if its mass is of the order of
$10^{-5}$ eV~\cite{darkmatter,kimrev,review}. It was also recently
argued~\cite{dundos} that an axion--photon oscillation can explain the observed
dimming supernovas~\cite{supern} if the axion has a rather small mass:
$\sim10^{-16}$ eV. Whatever of these possibilities (if any) is realized in
Nature, the existence of a light invisible axion can be spoiled by
gravity, since it induces renormalizable and non-renormalizable effective
interactions which break explicitly any global symmetry, in particular the
$U(1)_{\rm PQ}$
symmetry, and the axion can gain a mass which is greater than the mass coming
from instanton effects~\cite{gravity}. This can be avoided if the dimension of
the effective operators is $d\stackrel{>}{\sim}12$. Hence, unless $d$ is high
enough, invisible axion models do not solve the strong CP problem in a natural
way. Here we will show how in two electroweak models the axion is protected
against gravity effects, however the strategy can be used in any electroweak
model. 

It was pointed out several years ago by Krauss and Wilczek that a local gauge
symmetry, say ${\cal U}(1)$, can masquerade as discrete symmetries, $Z_N\subset
{\cal U}(1)$, to an observer equipped with only low-energy probes~\cite{kw}. It
means that these symmetries evade the no-hair theorem~\cite{nohair} i.e, unlike
continuous symmetries they must be respected even for gravitational
interactions. The only implication of the original gauge symmetry for the low
energy effective theory is the absence of interaction terms forbidden by the
$Z_N$ symmetry. For instance if there were more charged scalar fields in the
theory, the discrete symmetry would forbid many couplings that were otherwise
possible.  

Here we will not worry about the origin of this $Z_N$ (or ${\cal U}(1)$)
symmetry~\cite{dine92,hill,gl,ma,chang}. For instance, it might be that the
$Z_N$ symmetries come from a fifth dimension as it was shown for the case of
$U(1)_Y$ in Ref.~\cite{hill}. We will simply assume that at very high energies
we have a model of the form ${\cal U}(1)\otimes G_{EW}$ where ${\cal U}(1)$ is a
local symmetry (maybe a subgroup of a larger symmetry) which is broken to $Z_N$
at a high energy scale and $G_{EW}$ is an electroweak model i.e, ${\cal
U}(1)\otimes G_{EW}\to Z_N\otimes G_{EW}$. We will use the existence of these
local $Z_N$ symmetries in order to protect the axion against gravitational
effects. We get this by enlarging, if necessary, the representation content of
the model, so that we can impose symmetries with $N\geq13$. 
In addition the $U(1)_{\rm PQ}$ symmetry (and under some conditions also the
$Z_N$ symmetry) is an automatic symmetry, in the sense that it is not imposed on
the Lagrangian but it is just the consequence of the particle content of the
model, its gauge invariance, renormalizability and Lorentz invariance.  

In this circumstances we show how a naturally light invisible axion (it is
almost singlet $\phi$ under the gauge symmetry), which is also protected against
effects of quantum gravity, can be obtained in the context
of electroweak models as follows. Effective operators like
$\phi^{N-1}/M^{(N-1)-4}_{\rm Pl}$ are automatically suppressed by the (local)
$Z_N$ symmetry. At the same time this symmetry makes the $U(1)_{\rm PQ}$
symmetry an automatic symmetry of the classical Lagrangian or, as in the 3-3-1
model considered here, both $Z_N$ and $U(1)_{\rm PQ}$ are already automatic
symmetries of the model under the conditions discussed is Sec.~\ref{sec:331m}. 
For instance, a $Z_{13}$ symmetry implies that the first non-forbidden operator
is of dimension thirteen and it implies a contribution to the axion mass
square of $(v_\phi)^{11}/M^{9}_{\rm Pl}\approx 10^{-21} {\rm eV}^2$ or
$10^{-11}m^2_a$, if $m_a\sim\Lambda^2_{\rm
QCD}/v_\phi\approx10^{-5}$ eV is the instantons induced mass (we have used
$ M_{\rm Pl}=10^{19}$ GeV and $v_\phi=10^{12}$ GeV). The naturalness
of the PQ solution to the $\theta$-strong problem is not spoiled since in this
case we have $\theta_{\rm eff}\propto v^N_\phi/M^{N-4}_{\rm
Pl}\Lambda^4_{\rm  QCD}$~\cite{peccei99} and it means that $\theta_{\rm
eff} \propto 10^{-11}$ for $N=13$. 
Recently we applied this strategy in the context of an extension
of the electroweak standard model~\cite{iaxionsm}, now we will apply this
procedure to a model with 3-3-1 symmetry.

Hence, we see that it is necessary to search for mo\-dels which have a 
representation content large enough to allow the implementation of a discrete
symmetry $Z_N$ with $N\geq13$. In the context of a $SU(2)_L\otimes U(1)_Y$ model
we have enlarged the representation content by adding several Higgs doublets,
right-handed neutrinos and the scalar singlet necessary to make the axion
invisible. Hence, it is possible to accommodate a $Z_{13}$ symmetry while
keeping a general mixing among fermions of the same charge~\cite{iaxionsm}.
Larger $Z_N$ symmetries are possible if we add more scalar doublets in such a
way as to generate appropriate texture of the fermionic mass matrices.
On the other hand we will show in this work that in a 3-3-1 model, the
minimal representation content plus right-handed neutrinos, admit enough
large discrete symmetries. In these models the addition of a singlet (or a
decuplet~\cite{pal}) is the reason for maintaining the axion invisible,
however, in the 3-3-1 model the axion picture is a mixture of the Dine et al
invisible axion~\cite{dine} and the Kim's heavy quark axion model~\cite{kim}.
Nevertheless, unlike the model of Ref.~\cite{kim}, the exotic quarks are already
present in the minimal version of the model.  

This paper is organized as follows. In Sec.~\ref{sec:sm} we review briefly the
new invisible axion in the context of an extension of the standard
model~\cite{iaxionsm}. In Sec.~\ref{sec:331m} we consider one 3-3-1 model. Our
conclusions and some phenomenological consequences are in the last section.

\section{An extension of the standard model}
\label{sec:sm}

Let us consider the invisible axion in an extension of the $SU(2)_L\otimes
U(1)_Y$ model.  The representation content is the following:
$Q_L=(u\,d)^T_L\sim({\bf2},1/3)$, $L_L=(\nu\,l)^T_L\sim({\bf1},-1)$ denote any
quark and lepton doublet; $u_R\sim(\bf{1},4/3)$, 
$d_R\sim({\bf1},-2/3)$, $l_R\sim({\bf1},-2)$, $\nu_R\sim({\bf1},0)$ are the
right-handed components; and we will assume that each charge
sector gain mass from a different scalar doublet~\cite{glashow} i.e., $\Phi_u$,
$\Phi_d$, $\Phi_l$ and $\Phi_\nu$ ge\-ne\-ra\-te Dirac masses for $u$-like,
$d$-like quarks, charged leptons and neutrinos, respectively (all of them of the
form $({\bf2},+1)=(\varphi^+,\,\varphi^0)^T$). We also add a neutral complex
singlet $\phi\sim({\bf1},0)$ as in Refs.~\cite{dine,kim}, a singly charged
singlet $h^+\sim({\bf1},+2)$, as in the Zee's model~\cite{azee} and finally, a
triplet $\vec{T}\sim({\bf3},+2)$ as in Ref.~\cite{chengli}. The introduction
of right-handed neutrinos seems a natural option in any electroweak model if
neutrinos are massive particles, as strongly suggested by solar~\cite{solarnus},
reactor~\cite{kamland} and atmospheric~\cite{atmosnus} neutrino data. 

Next, we will impose the following (local in the sense discussed above) $Z_{13}$
symmetry among those fields:  
\begin{eqnarray}
Q_L\to \omega_5Q_L,\;u_R \to \omega_3u_R,\;
d_R\to \omega^{-1}_5 d_R,\nonumber \\ L\to \omega_6 L,\;
\nu_R\to \omega_0\nu_R,\; l_R\to \omega_4 l_R,\nonumber \\ 
\Phi_u\to \omega^{-1}_2\Phi_u,\; \Phi_d\to \omega^{-1}_3\Phi_d,\;
\Phi_l\to \omega_2\Phi_l,\nonumber \\
\Phi_\nu\to \omega^{-1}_6\Phi_\nu,\;
\phi\to\omega^{-1}_1\phi,\; \vec{T}\to w^{-1}_4\vec{T},\nonumber \\
 h^+\to \omega_1h^+
\label{z13}
\end{eqnarray}

with $\omega_k=e^{2\pi ik/13},\;k=0,1,...,6$. With this representation content
and the $Z_{13}$ symmetry defined in Eq.~(\ref{z13}) the allowed Yukawa
interactions and the scalar potential is automatically invariant under a
$U(1)_{\rm PQ}$ chiral symmetry. The PQ charges are quantized after imposing an
extra $Z_3$ symmetry with parameters denoted by $\tilde{\omega}_0$,
$\tilde{\omega}_1$ and $\tilde{\omega}^{-1}_1$. Under $Z_3$ the fields transform
as follows: 
\begin{eqnarray}
\Phi_u,\Phi_l,T,\nu_R\to \tilde{\omega}_1( \Phi_u,\Phi_l,T,\nu_R),\nonumber \\
\Phi_\nu,\phi,u_R,l_R\to \tilde{\omega}^{-1}_1(\Phi_\nu,\phi,u_R,l_R ),
\label{z3}
\end{eqnarray}
while all the other fields remain invariant i.e., transform with
$\tilde{\omega}_0$. As  we said before it is possible to implement larger $Z_N$
symmetries if more scalar doublets are added in such a way as to generate
appropriate texture of the fermionic mass matrices. 

The PQ assignment is the following: 

\begin{equation}
\begin{tabular}{ll}
$u'_L= e^{i(2/5) \alpha X_d}u_L$,& $d'_L= e^{-i\alpha X_d} d_L$,\\
$l'_L= e^{-i(4/5)\alpha X_d}l_L$,&
$\nu'_L= e^{i(3/5)\alpha  X_d}\nu_L$,\\
$\phi^{0\prime}_u= e^{-i(4/5)\alpha X_d}\phi^0_u$,&
$\phi^{0\prime}_d=  e^{-2i\alpha X_d}\phi^0_d$,\\
$\phi^{0\prime}_l=e^{-i(8/5)\alpha X_d}\phi^0_l$,&
$\phi^{0\prime}_\nu=  e^{-i(6/5)\alpha X_d}\phi^0_\nu$,\\
$\phi^{+\prime}_u= e^{i(3/5)\alpha X_d}\phi^+_u$,&
$ \phi^{+\prime}_d = e^{-i(3/5)\alpha X_d}\phi^+_d$,\\
$\phi^{+\prime}_l= e^{-i(8/5)\alpha X_d}\phi^+_l$,&
$\phi^{+\prime}_\nu= e^{i(1/5)\alpha X_d} \phi^+_\nu$, \nonumber \\
$T^{0\prime}= e^{-i(8/5)\alpha X_d}T^0$,&
$T^{+\prime}=e^{i(1/5)\alpha X_d}T^+$, \nonumber \\
$T^{++\prime}= e^{i(6/5)\alpha X_d}T^{++}$,& $h^{+\prime}=
e^{i(1/5)\alpha X_d}h^+$,\nonumber \\
$\phi'= e^{-i(6/5)\alpha X_d}\phi$ & \\
\label{pq1}
\end{tabular}
\end{equation}
The axion is invisible since it is almost singlet as in Ref.~\cite{dine,kim},
the scalar triplet is only a small correction for it.  

Some axion models~\cite{sikivie} lead to the formation of domain walls in the
evolution of the universe which could be inconsistent with the standard
cosmology~\cite{zeldovich}. The domain wall number is defined as~\cite{gw}
\begin{equation}
N_{DW}=\left\vert\sum_{f=L}{\rm Tr}X_fT^2_a(f)-\sum_{f=R}{\rm Tr}X_fT^2_a(f)
\large\right\vert,
\label{dw}
\end{equation}
where $X_f$ denotes the PQ charge of the quark $f$ and there is no summation
on $a$, we have ${\rm Tr}\,T_aT_b=(1/2)\delta_{ab}$ for ${\bf3}$ and ${\bf3}^*$.
Using the PQ assignment in Eq.~(\ref{pq1}) we obtain
$N_{DW}=\vert(9/5)X_d\vert$. If $X_d=\pm5$ we have $N_{\rm DW}=9$. (We choose 
$X_d=-5$ in order to have $\bar\theta\to\bar\theta+2\pi k,\,k=0,\cdots8$.) 
The $N=9$ vacua can be characterized by
\begin{eqnarray}
\langle u^\dagger_Lu_R\rangle=\mu^3_u\,e^{-i2\pi (2k/9)},\;
\langle d^\dagger_Ld_R\rangle=\mu^3_d\,e^{i2\pi (5k/9)},\nonumber \\
\langle\phi^0_u\rangle=v_u\,e^{i(\beta_u-2\pi(2k/9))},\;
\langle\phi^0_d\rangle=v_d\,e^{i(\beta_d-2\pi(5k/9))},\nonumber \\
\langle\phi^0_l\rangle=v_l\,e^{i(\beta_l-2\pi(4k/9))},\;
\langle\phi^0_\nu\rangle=v_\nu\, e^{i(\beta_\nu-2\pi(3k/9))},\nonumber \\
\langle T^0\rangle=v_T\,e^{i(\beta_T-2\pi(4k/9))},\;
\langle\phi\rangle=v_\phi\, e^{i(\beta_\phi-2\pi(3k/9))},
\label{vev}
\end{eqnarray}
where $k=0,\cdots,8$.

In this extension of the standard electroweak model only the $U(1)_{\rm PQ}$ 
is and automatic symmetry and its charges are quantized only if we add an extra
$Z_3$ symmetry. As we showed before there are genuine discrete symmetries that
are not broken by the instanton effects so, this model suffers of the domain
wall problem and it must be solved by any of the way proposed in
literature~\cite{gw,dw}. Moreover,  in this model the $Z_{13}$ symmetry is not
automatic. For more details see Ref.~\cite{iaxionsm}.  

\section{The axion in a 3-3-1 model}
\label{sec:331m}

The so called  3-3-1 models are interesting candidates for the physics at the
TeV scale~\cite{331,pt,mpp,dimo}. In fact, some years ago
Pal~\cite{pal} pointed out that the strong-CP question is solved elegantly
in those models. The point was that Yukawa couplings of these models
automatically contain a Peccei-Quinn symmetry~\cite{pq} if a simple discrete
symmetry was also imposed in order to avoid a trilinear term in the scalar
potential. Here we will consider one particular 3-3-1 model in which only three
scalar triplets are needed~\cite{pt} but we introduce also an scalar singlet,
$\phi\sim({\bf1},{\bf1},0)$. In this 3-3-1 model if we
add, {\it i)} right-handed neutrinos, {\it ii)} the conservation of the total
lepton number $L$ and, {\it iii)} a $Z_2$ symmetry defined below, we have that
both $Z_{13}$ and $U(1)_{\rm PQ}$ are accidental symmetries of the
classical Lagrangian (in the sense discussed in Sec.~\ref{sec:intro}).   

Before considering the implementation of a naturally light and invisible
axion in the context of the 3-3-1 model of Ref.~\cite{pt}, let us briefly review
the model.  

In the quark sector we have: 
\begin{eqnarray}
Q_{mL}=(d_m,\, u_m,\, j_m)^T_L\sim ( {\bf3}, {\bf3}^{*},- 1/3),\;
m=1,2;\nonumber \\
Q_{3L}=(u_3,\, d_3,\,J)^T_L\sim ( {\bf3}, {\bf 3}, 2/3),
\label{quarksl}
\end{eqnarray}
and the respective right-handed components in singlets
\begin{eqnarray}
u_{\alpha R}\sim({\bf3},{\bf1},2/3), d_{\alpha R}\sim({\bf3},{\bf1},-1/3),\,
\alpha=1,2,3;\nonumber \\
 J_{R}\sim({\bf3},{\bf1},5/3);\,j_{mR}\sim({\bf3},{\bf1},-4/3).
\label{quarksr}
\end{eqnarray}

In the scalar sector, this model has only three triplets 
\begin{eqnarray}
\chi=(\chi^{-},\,\chi ^{--},\, \chi^0)^T,\, 
\rho =(\rho^{+},\,\rho ^0,\,\rho ^{++})^T ,\nonumber
\\ \eta =(\eta^0,\,\eta _1^{-},\, \eta _2^{+})^T,
\label{triplets}
\end{eqnarray}
transforming as $({\bf1}, {\bf 3},-1), ( {\bf1}, {\bf 3}, 1)$ and 
$({\bf1},{\bf3},0)$, respectively. Finally, in this model leptons transform as
triplets $({\bf 3}_a,0)$ where $a=e,\mu,\tau$:
\begin{equation}
\Psi_{aL}=(\nu_a,\,l_a,\,E_a)^T_L,
\label{mb}
\end{equation}
and the respective right-handed singlets
\begin{equation}
\nu_{aR}\sim({\bf1},{\bf1},0),\;l_{aR}\sim({\bf1},{\bf1},-1),\;E_{aR}\sim
({\bf1},{\bf1},+1),
\label{mbr}
\end{equation}
and we have added right-handed neutrinos which are not present in the minimal
version of the model.
Hence, we see that the model has fifteen multiplets, including right-handed
neutrinos and the singlet $\phi$ (in fact, as usual we have to add a scalar
singlet $\phi\sim({\bf1},{\bf1},0)$ in order to make the axion
invisible~\cite{dine,kim}) and it will admit, under the three condition
introduced above, automatic $Z_{13}$ and $U(1)_{\rm PQ}$ symmetries as we will
show in the following. 

With the quark and scalar multiplets above we have the Yukawa
interactions
\begin{eqnarray}
-{\cal L}^q_Y&=& 
\overline{Q}_{iL} ( F_{i\alpha }u_{\alpha
R}\rho ^{*}+\widetilde{F}_{i\alpha }d_{\alpha R}\eta ^{*}) 
+ \overline{Q}_{3L} ( G_{1\alpha }
u_{\alpha R}\eta \nonumber \\ &+&
\widetilde{G}_{1\alpha }d_{\alpha R}\rho) + 
\lambda _1\overline{Q}_{3L}J_{1R}^{\prime }\chi+
\lambda _{im}\overline{Q}_{iL}j_{mR}\chi ^{*}\nonumber \\ &+&H.c.,
\label{yu1}
\end{eqnarray}
where repeated indices mean summation. In the lepton sector we have 
\begin{eqnarray}
-{\cal L}^l_Y&=&G^\nu_{ab}\overline{(\Psi)}_{aL}
\nu_{bR}\,\eta
+G^l_{ab}\overline{(\Psi) }_{aL}l_{bR}\rho+
G^E_{ab}\overline{(\Psi) }_{aL}E_{bR}\chi \nonumber \\ &+&H.c. 
\label{yu2} 
\end{eqnarray}

In both Yukawa interactions above a general mixing in each
charge sector is allowed. If we want to implement a given texture for the quarks
and lepton mass matrices we have to introduce more scalar triplets and a larger
$Z_N$ symmetry will be possible in the model. Interesting possibilities are the
cases where $N$ is a prime number (see below). 

The most general $L-$, $Z_2-$ and gauge invariant scalar potential is   
\end{multicols}
\hspace{-0.5cm}
\rule{8.7cm}{0.1mm}\rule{0.1mm}{2mm}
\widetext
\begin{eqnarray}
V_{\rm 331} & = & \sum_{x=\eta,\rho,\chi,\phi}\mu^2_x T^\dagger_xT_x+
\lambda_1(\eta^\dagger \eta)^2+ \lambda_2(\rho^\dagger \rho)^2+
\lambda_3(\chi^\dagger \chi)^2 +
\left(\eta^\dagger\eta\right)\left[\lambda_4\left(\rho^\dagger 
\rho\right) + \lambda_5\left(\chi^\dagger\chi\right)\right]\nonumber \\& +&
 \lambda_6 \left(\rho^\dagger\rho\right)\left(\chi^\dagger\chi\right) +  
 \lambda_7\left(\rho^\dagger\eta\right)\left(\eta^\dagger\rho\right)
+\lambda_8\left(\chi^\dagger\eta\right)\left(\eta^\dagger\chi\right)+ 
 \lambda_9\left(\rho^\dagger\chi\right)\left(\chi^\dagger\rho\right)
 \nonumber \\ &+&
\lambda_\phi(\phi^*\phi)^2+ \phi^*\phi\sum_{k=\eta,\rho,\chi}
\lambda_{\phi k}T^\dagger_kT_k+
\left(\lambda_{10}\,\phi\,\epsilon^{ijk}\eta_i\rho_j\chi_k + \mbox{H. c.}\right),
\label{pe1}
\end{eqnarray}

\hspace{9.1cm}
\rule{-2mm}{0.1mm}\rule{8.7cm}{0.1mm}
\begin{multicols}{2}
\narrowtext
\noindent where $\mu'$s are parameters with dimension of mass, $\lambda'$s are
dimensionless parameters and we have denoted $T_x=\eta,\rho,\chi,\phi$ and $T_k=
\eta,\rho,\chi$. Now we can explain the motivation of the three conditions
assumed at the beginning of the section: {\it i)} with the present experimental
data~\cite{solarnus,kamland,atmosnus} in any electroweak model
right-handed neutrinos are no longer avoided under the assumption that neutrinos
are massless; {\it ii)} in Eq.~(\ref{yu1}) it is still possible a Majorana mass term 
among right-handed neutrinos, say $M_R\overline{(\nu^c)_R}\nu_R$; on the other
hand in the scalar potential it is possible to have the quartic term 
$\chi^\dagger\eta\rho^\dagger\eta$ which also violates the total lepton number.
Both terms are avoided by imposing the conservation of the total lepton number 
$L$; {\it iii)} the trilinear term in the scalar potential $\eta\rho\chi$ is
avoided if we impose a $Z_2$ symmetry under which $J_R,j_{mR},\chi,\phi$ are odd
and all the other fields are even. In this conditions, the Yukawa interactions
in Eqs.(\ref{yu1}), (\ref{yu2}) and the scalar potential in Eq.~(\ref{pe1} ) are
automatically invariant under the (local) $Z_{13}$ symmetry:   

\begin{eqnarray}
Q_{iL}\to \omega^{-1}_5Q_{iL},\;\;Q_{3L}\to\omega_5Q_{3L},\nonumber \\
u_{\alpha R} \to \omega_1u_{\alpha R},\;\;
d_{\alpha R}\to \omega^{-1}_1 d_{\alpha R},\nonumber \\
J_R\to \omega_3 J_R,\;\;j_{mR}\to\omega^{-1}_3 j_{mR},\nonumber \\
\Psi_L\to w_0\Psi_L,\;\; l_R\to\omega^{-1}_6l_R,\nonumber \\ 
\nu_R\to\omega^{-1}_4\nu_R,\;\;E_R\to\omega^{-1}_2E_R,\nonumber \\
\eta\to \omega_4\eta,\;\; \rho\to \omega_6\rho,\nonumber \\
\chi\to \omega_2\chi,\;\;\phi\to\omega_1\phi,
\label{z13b}
\end{eqnarray}
where $\omega_k=e^{2\pi ik/13},\,k=0,\cdots,6$. Notice that if $N$ is a
prime number the singlet $\phi$ can transform under this symmetry with any
assignment (but the trivial one), o\-ther\-wi\-se we have to be careful with the
way we choose the singlet $\phi$ transforms under the $Z_N$ symmetry. This
symmetry implies that the lowest order effective operator which contributes to
the axion mass is $\phi^{13}/M^9_{\rm Pl}$ which gives a mass of the order
$(v_\phi)^{11}/M^9_{\rm Pl}$ and also keeps the $\bar{\theta}$ parameter small 
as discussed previously.  

It happens that as the $Z_{13}$ symmetry, the $U(1)_{\rm PQ}$ is also automatic
i.e., a consequence of the gauge symmetry and renormalizability of the model, in
the interactions in Eqs.~(\ref{yu1}), (\ref{yu2}) and (\ref{pe1}).
Let us see the PQ charge assignment for the fermions in the model:
\begin{eqnarray}
u'_L= e^{-i\alpha X_u}u_L,\; d'_L= e^{-i\alpha X_d}d_L,\;\;
l'_L= e^{-i\alpha X_l}l_L,\nonumber \\
\nu'_L=e^{-i\alpha X_\nu}\nu_L,\;
j'_L=e^{-i\alpha X_j}j_L,\;
J'_L= e^{-i\alpha X_J}J_L,\; \nonumber \\
E'_L=e^{-i\alpha X_E}E_L,\;
\label{pq2}
\end{eqnarray}
and in the scalar sector we have the following PQ charges:
\begin{eqnarray}
\eta^{\prime0}= e^{-2i\alpha X_u}\eta^0 =  e^{+2i\alpha X_d}\eta^0,\nonumber \\
\eta^{\prime-}_1= e^{-i\alpha(X_u+X_d)} \eta^-_1=
e^{+i(X_u+X_d)}\eta^-_1,\nonumber \\
\eta^{+\prime}_2=e^{-i\alpha(X_J+X_u)}\eta^+_2= e^{+i\alpha(X_j+X_d)}\eta^+_2,
\nonumber \\
\rho^{\prime0}=  e^{+2i\alpha X_u}\rho^0= e^{-2i\alpha X_d}\rho^0 ,\nonumber \\
\rho^{\prime+}= e^{-i\alpha(X_u+X_d)}\rho^+=e^{+i\alpha(X_u+X_d)}\rho^+,
\nonumber \\
\rho^{\prime++}= e^{-i\alpha(X_J+X_d)}\rho^{++}=e^{+i\alpha(X_j+X_u)}\rho^{++},
\nonumber \\
\chi^{\prime-}=e^{-i\alpha(X_u+X_J)}\chi^-=
e^{+i\alpha(X_d+X_j)}\chi^-,\nonumber \\
\chi^{\prime--}= e^{-i\alpha(X_d+X_J)}\chi^{--}=
e^{+i\alpha(X_u+X_j)}\chi^{--},\nonumber \\ 
\chi^{\prime0}= e^{-2i\alpha X_J} \chi^0= e^{+2i\alpha X_j} \chi^0,\nonumber \\
 \phi'= e^{-2i X_j}\phi. 
\label{pq3}
\end{eqnarray}
From Eqs.~(\ref{pq2}) and (\ref{pq3}) we obtain the following relations
\begin{equation}
X_d=-X_u=X_l=-X_\nu,\quad  X_j=-X_J=-X_E. 
\label{2charges}
\end{equation}
In the present model, although we
have two independent PQ charges (say, $X_d$ and $X_j$), the known quark
contributions to $\bar\theta$ which are proportional to $X_d$ cancel out
exactly. Only $X_j$ is important for solving the $CP$ problem:
\begin{equation}
\bar\theta\to \bar\theta-2\alpha\sum_{{\rm all}\,{\rm 
quarks}}X_f =\bar{\theta}-2\alpha X_j.    
\label{tb}
\end{equation}

Hence we can assume that $X_d=0$ and the only relevant PQ
transformations are 
\begin{eqnarray}
j'_L=e^{-i\alpha X_j}j_L,\;
J'_L= e^{i\alpha X_j}J_L,\; \nonumber \\
E'_L=e^{-i\alpha X_j}E_L,\;
\eta^{+\prime}_2=e^{i\alpha X_j}\eta^+_2,
\nonumber \\
\rho^{\prime++}= e^{i\alpha X_j}\rho^{++},\;
\chi^{\prime-}=e^{i\alpha X_j}\chi^- \nonumber \\
\chi^{\prime--}= e^{i\alpha X_j}\chi^{--},\; 
\chi^{\prime0}= e^{2i\alpha X_j} \chi^0,\nonumber \\
 \phi'= e^{-2i\alpha X_j}\phi. 
\label{pq4}
\end{eqnarray}

Notice that like in Ref.~\cite{kim} we have an invisible a\-xi\-on (it is almost 
singlet, see below) and the PQ charge which solves the strong CP problem is
the charge of the e\-xo\-tic quarks. However, unlike Ref.~\cite{kim} the
heavy quarks is already present in the minimal version of this 3-3-1 model.
The condition $X_u=-X_d$ is not allowed in the context
of the standard $SU(2)_L\otimes U(1)_Y$ model since in this case it is not
possible to shift the $\overline{\theta}$. 

We have seen above that the known quarks contributions to $\bar\theta$ 
cancel out exactly. This also happens in the domain wall number, $N_{\rm DW}$,
defined in Eq.~(\ref{dw}). Using the PQ charges in Eq~(\ref{pq2}) and
(\ref{pq3}) we obtain $N_{DW}=X_j$ and since we have always chosen $X_j=1$ we
see that as in the Kim's axion model, there is no domain wall problem in this
3-3-1 axion model. We stress that the contribution of the known quarks cancel
out exactly even if we assume that $X_d\not=0$. Moreover, we will see in
Sec.~\ref{sec:con} that also the coupling of the axion with photon does not
depend at all on the PQ charge of the usual quarks and
leptons since there is an exact cancellation among them. However if $X_d\not=0$
there are still couplings with the usual fermions. Notwithstanding since $X_d$
does not play any role in the solution of the strong $CP$ problem we can assume
at the very start that $X_d=0$ i.e., that the nontrivial PQ transformations are
those in Eq.~(\ref{pq4}). It means that at the tree level there is no coupling
of the axion with the known quarks and charged leptons.  

We can verified that in fact the axion is almost singlet. After redefining the
neutral fields as usual $T^0_k=(v_k+{\rm Re}T^0_k+i{\rm
Im}T^0_k)/\sqrt2,\,\phi=(v_\phi+{\rm Re}\phi+i{\rm Im}\phi)/\sqrt2$, 
with $k=\eta,\rho,\chi$, we obtain the constraint
equations $t_x=0$, (where $x=\eta,\rho,\chi,\phi$): 
\end{multicols}
\hspace{-0.5cm}
\rule{8.7cm}{0.1mm}\rule{0.1mm}{2mm}
\widetext
\begin{eqnarray}
&&t_\eta={\rm Re}[\mu^2_\eta v_\eta+\lambda_1\vert v_\eta\vert^2 v_\eta+
\frac{1}{2}\,(\lambda_4 \vert v_\rho\vert^2+\lambda_5\vert v_\chi\vert^2)v_\eta+
\frac{\lambda_{10}}{2}v_\rho v_\chi v_\phi+\lambda_{\phi\eta} \vert
v_\phi\vert^2 v_\eta],
\nonumber \\&& t_\rho={\rm Re}[\mu^2_\rho v_\rho+\lambda_2 \vert v_\rho\vert^2
v_\rho+ \frac{1}{2}\,(\lambda_4 \vert v_\eta\vert^2+\lambda_6 \vert v_\chi\vert^2)v_\rho+
\frac{\lambda_{10}}{2}v_\eta v_\chi v_\phi+\lambda_{\phi\rho} \vert
v_\phi\vert^2 v_\rho],
\nonumber \\ && 
t_\chi={\rm Re}[\mu^2_\chi v_\chi+\lambda_3 \vert v_\chi\vert^2v_\chi
+\frac{1}{2}(\lambda_5 \vert v_\eta\vert^2+
\lambda_6 \vert v_\rho\vert^2)v_\chi+\frac{\lambda_{10}}{2}v_\eta v_\rho v_\phi+
\lambda_{\phi\chi} \vert v_\phi\vert^2 v_\chi], \nonumber \\ &&
t_\phi={\rm Re}[\mu^2_\phi v_\phi+\lambda_\phi \vert v_\phi\vert^2v_\phi +
\frac{\lambda_{10}}{2} v_\eta v_\rho v_\chi+
\frac{1}{2}(\lambda_{\phi\eta}\vert v_\eta\vert^2+
\lambda_{\phi\rho}\vert v_\rho\vert^2+\lambda_{\phi\chi}
\vert v_\chi\vert^2)v_\phi],\nonumber \\ && {\rm Im}(v_\phi v_\eta v_\rho
v_\chi)=0. 
\label{ce}
\end{eqnarray}
\hspace{9.1cm}
\rule{-2mm}{0.1mm}\rule{8.7cm}{0.1mm}
\begin{multicols}{2}
\narrowtext
\noindent Notice that the solution with $v_\eta,v_\rho,v_\chi,v_\phi\not=0$
is allowed. For instance, just for an illustration, assuming for simplicity that
all VEVS and parameters (but $\mu_\phi$ and $\mu_\chi$) 
are real, $\lambda_1v^2_\eta,\lambda_2 v^2_\rho,\vert\lambda_{\phi k}
v_kv_\phi\vert\ll\vert\lambda_{10}v_kv_{k'}\vert$, we obtain 
\begin{eqnarray}
v^2_\phi\approx-\frac{\mu^2_\phi}{\lambda_\phi},\;
v^2_\chi\approx-\frac{\mu^2_\chi}{\lambda_3}\nonumber \\
v_\eta\approx-\lambda_{10}\frac{v_\rho v_\chi v_\phi}{\mu^2_\eta},\;
v_\rho\approx-\lambda_{10}\frac{v_\eta v_\chi v_\phi}{\mu^2_\rho},
\label{vevs2}
\end{eqnarray}
and the self-consistent condition
$\lambda^2_{10}v^2_\chi v^2_\phi\approx\mu^2_\eta\mu^2_\rho$.
With all VEVs real the pseudoscalar mass eigenstates, in the basis ${\rm
Im}(\eta^0,\rho^0,\chi^0,\phi$), are given by:  
\begin{eqnarray}
G^0_1=\frac{1}{\left(v^2_\eta+v^2_\rho\right)^{1/2}}
\left(-v_\eta,v_\rho,0,0\right),\nonumber \\
G^0_2=\frac{1}{\sqrt{N}}\left(\frac{s_1}{s_2} v^2_\rho  v_\eta,
\frac{s_1}{s_2} v^2_\eta
v_\rho,\frac{s_2}{s_1} v^2_\phi v_\chi, \frac{s_2}{s_1} 
v^2_\chi v_\phi\right),\nonumber \\
a=\frac{1}{\left(v^2_\chi+v^2_\phi\right)^{1/2}}
\left(0,0,-v_\chi,v_\phi\right),\nonumber \\
A^0=\frac{1}{\sqrt{N}}\left(v_\rho v_\chi v_\phi,v_\eta v_\chi v_\phi, v_\eta
v_\rho v_\phi, v_\eta v_\rho v_\chi\right).
\label{autovetores}
\end{eqnarray}
where $s_1=(v^2_\chi+ v^2_\phi)^{1/2}$,
$s_2=(v^2_\eta+v^2_\rho)^{1/2}$ and $
N=v^2_\eta\,[\,v^2_\rho(v^2_\chi +v^2_\phi)+v^2_\chi v^2_\phi]+
v^2_\rho v^2_\chi v^2_\phi$;
$G^0_{1,2}$ are genuine Goldstone bosons (eaten when the gauge bosons
$Z,Z^\prime$ become massive), $a$ is the axion pseudo-Goldstone.
If $\vert v_\phi\vert\gg \vert v_k\vert$ we see that
$a\simeq {\rm Im }\phi$ and we have an invisible axion~\cite{dine}. The
usual restriction coming from red giant implies $\vert v_\phi\vert>10^9$
GeV~\cite{redg};  $A^0$ is a heavy pseudo-scalar with 
$m^2_A=-(\lambda_{10}/8C)[v^2_\eta[v^2_\rho (v^2_\chi+ v^2_\phi)
+v^2_\chi v^2_\phi] + v^2_\rho v^2_\chi v^2_\phi]$, with $\lambda_{10}<0$
and we have defined $C=v_\phi v_\eta v_\rho v_\chi$. 
 
\section{conclusion}
\label{sec:con}

We have build invisible axion models in which the axion is
naturally light (protected against quantum gravity effects) because of a
$Z_{13}$ discrete symmetry. In the context of a $SU(2)\otimes U(1)$ model
this symmetry and a $Z_3$ have to be imposed and new fields have to be added. 
On the other hand, in a 3-3-1 model with right-handed neutrinos added, the
$Z_{13}$ is automatic if we impose the conservation of the total lepton number
and a $Z_2$ symmetry. Moreover, in both models $U(1)_{\rm PQ}$ 
is an accidental symmetry. It means that at low energy the gauge
symmetries are $SU(3)_C\otimes SU(2)_L\otimes U(1)_Y\otimes Z_{13}\otimes Z_3$
or $SU(3)_C\otimes SU(3)_L\otimes U(1)_Y\otimes Z_{13}\otimes Z_2$. Notice
however that in the context of the standard model even by imposing $L$
conservation and the $Z_3$ symmetry, the $Z_{13}$ symmetry is not automatic but
in the 3-3-1 model $L$ conservation and $Z_2$ make $Z_{13}$ an automatic
symmetry.  

Hence, we have implemented an invisible and naturally light axion in a multi
Higgs extension of the standard model and also in a 3-3-1
model. Unlike the axion of the first model, in
the 3-3-1 model considered here the minimal representation content (plus
right-handed neutrinos) is already enough to implement a local $Z_{13}$
symmetry. As we said before, the interesting
discrete symmetries are those in which $N$ is a prime number. In this case $Z_N$
has no subgroup except itself and the identity~\cite{hall}. Hence, the next
interesting symmetry should be $Z_{17}$ which will allow even lighter
axion, i.e., $m^2_{a{\rm (gravity)}}= (v_\phi)^{15}/M^{13}_{\rm PL}\sim 10^{-67}
\,{\rm eV}^2$ or $10^{-57}m^2_a$, similar we have in this case
$\bar\theta_{\rm (gravity)}\sim10^{-43}$. 

We will consider now some phenomenological consequences of the axion in the
3-3-1 model. (The case of the model of Sec.~\ref{sec:sm} has been considered in
Ref.~\cite{iaxionsm}.) In general the axion-photon coupling is given by
\begin{equation}
c_{a\gamma\gamma}=\tilde{c}_{a\gamma\gamma}-1.95,
\label{agg}
\end{equation}
where the first term is defined as
\begin{equation}
\tilde{c}_{a\gamma\gamma} =\frac{1}{N_{DW}}\sum_{{\rm all}\,{\rm 
fermions}}X_f\,Q^2_f,    
\label{aggt}
\end{equation}
with $N_{DW}=1$ in this model since it has no domain wall
problem as in the Kim's model~\cite{kim}; $X_f$ and $Q_f$ are
the PQ and electromagnetic charge, respectively, of the fermion $f$. The term
$-1.95$ comes from the light quark PQ anomalies and it does exist only if these
quarks carry PQ charges. In Eq.~(\ref{aggt}) the contribution proportional to
$X_d$ cancels out exactly even if we had assumed that $X_d\not=0$. So we have in
general that $c_{a\gamma\gamma}=-(2/3)X_j-1.95$, or
$c_{a\gamma\gamma}=2.62(-1.28)$ for $X_j=+1(-1)$. In our case in particular,
since $X_d=0$, we have just the first contribution in Eq.~(\ref{agg}) i.e.,
$c_{a\gamma\gamma}=\tilde{c}_{a\gamma\gamma} =\pm(2/3)$. We stress that the
contributions of the PQ charges of the known quarks cancel out exactly in
$\bar{\theta}$ in Eq.~(\ref{tb}), in the domain wall number given in
Eq.~(\ref{dw}), and also in Eq.~(\ref{aggt}). Thus, if we assign PQ charges to
those quarks it only implies in a coupling with the axion at the tree level. On
the other hand, if we assume that only the exotic fermions of the model carry PQ
charge the coupling with the known quarks and leptons arises at higher order in
perturbation theory. 
 
We have imposed at the very start the conservation of the total lepton number
and a $Z_2$ symmetry. Another possibility is to impose the
discrete symmetry $Z_{13}$. In this case only the $U(1)_{\rm PQ}$ symmetry is
automatic and the quartic $L$-violating term
$\chi^\dagger\eta\rho^\dagger\eta$ is allowed. This term implies 
a new relation among the PQ charges of the known particles and the exotic ones:
$X_j=3X_d$. Taking also into account Eqs.~(\ref{2charges}), it can be shown
that there is a surviving symmetry $Z_3\subset U(1)_{\rm PQ}$ which
implies a domain wall problem~\cite{gl2}.   

The 3-3-1 model in which the leptons transform as
$\Psi_{aL}=(\nu_a,\,l_a,\,l^c_a)^T_L$ needs also the introduction of a scalar
sextet $S\sim({\bf6},0)$ (or singlet charged leptons
$E\sim({\bf1},1)$)~\cite{dma,lepmass1}.
In this case the Yukawa interaction in the quark sector is given by 
Eq.~(\ref{yu1}) and  in the leptonic sector we have
\begin{eqnarray}
-{\cal L}^l_Y&=&G^\nu_{ab}\overline{\Psi }_{aL}
\nu_{bR}\,\eta
+G_{ab}\overline{(\Psi)^c }_{aL}\psi_{bL}S \nonumber \\ &+&
G^\prime_{ab}\epsilon_{ijk}
\overline{(\Psi)^c}_{iaL}\psi_{jbL}\eta_k 
+H.c. 
\label{yu3}
\end{eqnarray}
where $G^\nu_{ab}$ is an arbitrary $3\times3$ matrix while
$G_{ab} (G^\prime)$ is a symmetric (antisymmetric) matrix and we have omitted
some $SU(3)$ indices. Notice that this model has only thirteen multiplets
(including right-handed neutrinos and the singlet $\phi$) so we
can not have a symmetry as large as $Z_{13}$. However by adding more scalar
multiplets like in Ref.~\cite{effop2} it may be possible to implement,
automatically, a large enough $Z_N$ symmetry. The 
supersymmetric version of the model can also be considered since in
this model, without considering right-handed neutrinos, there are 23 chiral
superfields (in the same case, the MSSM has 14 chiral superfields)~\cite{331s}. 

In the models considered in this work the axion couples to neutrinos too.
This coupling may have astrophysics and/or cosmological consequences;
we can also implement hard~\cite{hard}, soft~\cite{cp3} or
spontaneous~\cite{laplata} CP violation.

\acknowledgments 
This work was supported by Funda\c{c}\~ao de Amparo \`a Pesquisa
do Estado de S\~ao Paulo (FAPESP), Conselho Nacional de 
Ci\^encia e Tecnologia (CNPq) and by Programa de Apoio a
N\'ucleos de Excel\^encia (PRONEX).

\end{multicols}

\end{document}